\documentclass[11pt,twoside]{article}

\usepackage{asp2006}
\usepackage{epsf}
\usepackage{psfig}

\begin{document}
\title{RS Ophiuchi in Quiescence: Why is it X-ray Faint?}   
\author{Koji Mukai}   
\affil{CRESST and X-ray Astrophysics Laboratory NASA/GSFC, Greenbelt, MD 20771 \\
and Department of Physics, University of Maryland, Baltimore County, 1000 Hilltop Circle, Baltimore, MD 21250}.     

\begin{abstract} 
The short interval between successive outbursts of RS~Oph strongly
suggests that it has a high mass white dwarf accreting at a high rate.
This, in turn, suggests the possibility of prominent X-ray emission
from RS~Oph in quiescence.  However, archival quiescent X-ray observations
of RS~Oph show it to be a modest soft X-ray source but not a strong
2--10 keV X-ray source.  In this aspect, RS~Oph differs markedly from
T~CrB.  We speculate on the possible mechanisms that could significantly
suppress the 2--10 keV X-ray emission in RS~Oph.
\end{abstract}


\section{Motivation}

The 2006 outburst of RS Oph is almost certainly due to a thermonuclear
runaway of hydrogen rich fuel that has accumulated since the previous
outburst in 1985.  According to the theories of nova outbursts, both
the white dwarf mass and the mass accretion rate must be high for
two consecutive nova outbursts to be separated by such a short interval.
According to the calculations of \cite{Yea2005}, a 1.25 M$_\odot$ white
dwarf accreting at $\sim$10$^{-7}$ M$_\odot$\,yr$^{-1}$ would accumulate a
sufficient amount of fuel in 20 years.  This corresponds to a boundary
layer luminosity of $\sim$1.4 $\times 10^{36}$ ergs\,s$^{-1}$, assuming
that half the potential energy is radiated away in the accretion disk
and the white dwarf is rotating slowly.  If RS~Oph harbors a 1.4 M$_\odot$
white dwarf, it requires less fuel to achieve a thermonuclear runaway.
Even then, an accretion rate of $\sim$10$^{-8}$ M$_\odot$\,yr$^{-1}$ is
required, implying a boundary layer luminosity of 
$\sim$3 $\times 10^{35}$ ergs\,s$^{-1}$.  The boundary layer emission
is likely to be in the hard or soft X-rays \citep{PR1985a,PR1985b},
where the contribution from the red giant mass donor is expected
to be minimal.

\section{Quiescent X-ray Observations}

We have therefore searched the HEASARC archive for past, pointed
X-ray observations of RS~Oph, excluding those that were performed during
the 1985 and 2006 outbursts, and found 7 observations using 4 satellites.
These quiescent X-ray observations are summarized in Table 1.  In
calculating the luminosity, the distance to RS~Oph was assumed to
be 1.44 kpc, and no correction was made for absorption.

\begin{table}[!ht]
\caption{Quiescent X-ray Observations of RS Oph}
\smallskip
\begin{center}
{\small
\begin{tabular}{cccccc}
\tableline
\noalign{\smallskip}
Date & Satellite & Sequence & Energy Range & Count Rate & Luminosity \\
     &           &          & (keV)        & (ct\,s$^{-1}$) & (ergs\,s$^{-1}$) \\
\noalign{\smallskip}
\tableline
\noalign{\smallskip}
1979-09-25 & {\sl Einstein} & I03190         & 0.2--3.5  & $< 6 \times 10^{-3}$ & $< 3.4 \times 10^{31}$ \\
1981-04-03 & {\sl Einstein} & I10703         & 0.2--3.5  & $< 2 \times 10^{-3}$ & $< 1.1 \times 10^{31}$ \\
1991-03-02 & {\sl ROSAT}    & RP300038A00    & 0.2--2.4  & 0.0035               & 9.4$\times 10^{30}$ \\
1992-03-04 & {\sl ROSAT}    & RP300038A01    & 0.2--2.4  & 0.0116               & 3.1$\times 10^{31}$ \\
1997-02-04 & {\sl RXTE}     & 20027-01-01-00 & 2--10     & 0.16                 & 2.4$\times 10^{32}$ \\
1997-02-06 & {\sl RXTE}     & 20027-01-01-01 & 2--10     & 0.10                 & 1.7$\times 10^{32}$ \\
1997-03-13 & {\sl ASCA}     & 35003000       & 0.7--2.0  & 1.3$\times 10^{-3}$  & 8.6$\times 10^{30}$ \\
           &                &                & 2--10     & 1.2$\times 10^{-3}$  & $\leq 5 \times 10^{31}$ \\
\noalign{\smallskip}
\tableline
\end{tabular}
}
\end{center}
\end{table}

RS~Oph was observed twice with {\sl Einstein} IPC well before the 1985
outburst, and was undetected both times.  The exposure times were 1.7
ksec in 1979 Sep and 5.5 ksec in 1981 Apr.  In Table 1, we give approximate
upper limits, based on an estimate that 10 source counts would have
resulted in a detection.  The luminosity limits are then inferred assuming
the plasma model parameters that fit the {\sl ROSAT\/} data described
below.

The {\sl ROSAT\/} PSPC detections have already been published by
\cite{O1993} and by \cite{Oea2001}.  The former reported on
the first observation, and the latter was a compilation of a large
number of {\sl ROSAT\/} observations of novae and recurrent novae.
In Figure~1 (left), we show the PSPC spectrum from the second
{\sl ROSAT\/} observation, together with a {\tt mekal} model fit,
from which we obtain a plasma temperature of kT$\sim$0.6 keV and
an absorbing column of N$_{\rm H} \sim$9.6 $\times 10^{21}$ cm$^{-2}$.
Note, however, that the fit is neither unique nor well constrained
due to the low statistical quality of the data.  The estimated N$_{\rm H}$
value suggests that the intrinsic (unabsorbed) luminosity is about
10 times higher than the observed (absorbed) value.  While these
{\sl ROSAT\/} observations do establish RS~Oph as a soft X-ray
source, what about the 2--10 keV band?

The two {\sl RXTE\/} observations should be considered non-detections,
even though the observed count rates are higher than the predictions of
the background model.  The count rate excesses are statistically significant,
and are above the estimated systematic errors of the background model
that reproduces the non X-ray background and the average cosmic X-ray
background.  However, give the location of RS~Oph and the collimator
response of {\sl RXTE\/} PCA, contamination due to Galactic X-ray
background and faint, discrete sources in the field of view at the
observed level is quite possible.  Imaging X-ray telescopes are necessary
to securely detect RS~Oph in quiescence.

The {\sl ASCA\/} observation with an exposure time of $\sim$34 ksec
is the most sensitive observation to date of 2--10 keV emission from
RS~Oph in quiescence.  Of the two types of detectors on-board {\sl ASCA\/},
SIS is more sensitive below 2 keV, while GIS is more sensitive for
harder emission.  The values tabulated in Table\,1 are from GIS in
two energy bands.  While RS~Oph is securely detected below 2 keV
($1.3 \pm 0.4 \times 10^{-3}$ cts\,s$^{-1}$), it is not so in the 2--10
keV band ($1.2 \pm 0.8 \times 10^{-3}$ cts\,s$^{-1}$).  The analysis
of SIS data produces similar results.

\begin{figure}[!ht]
\plottwo{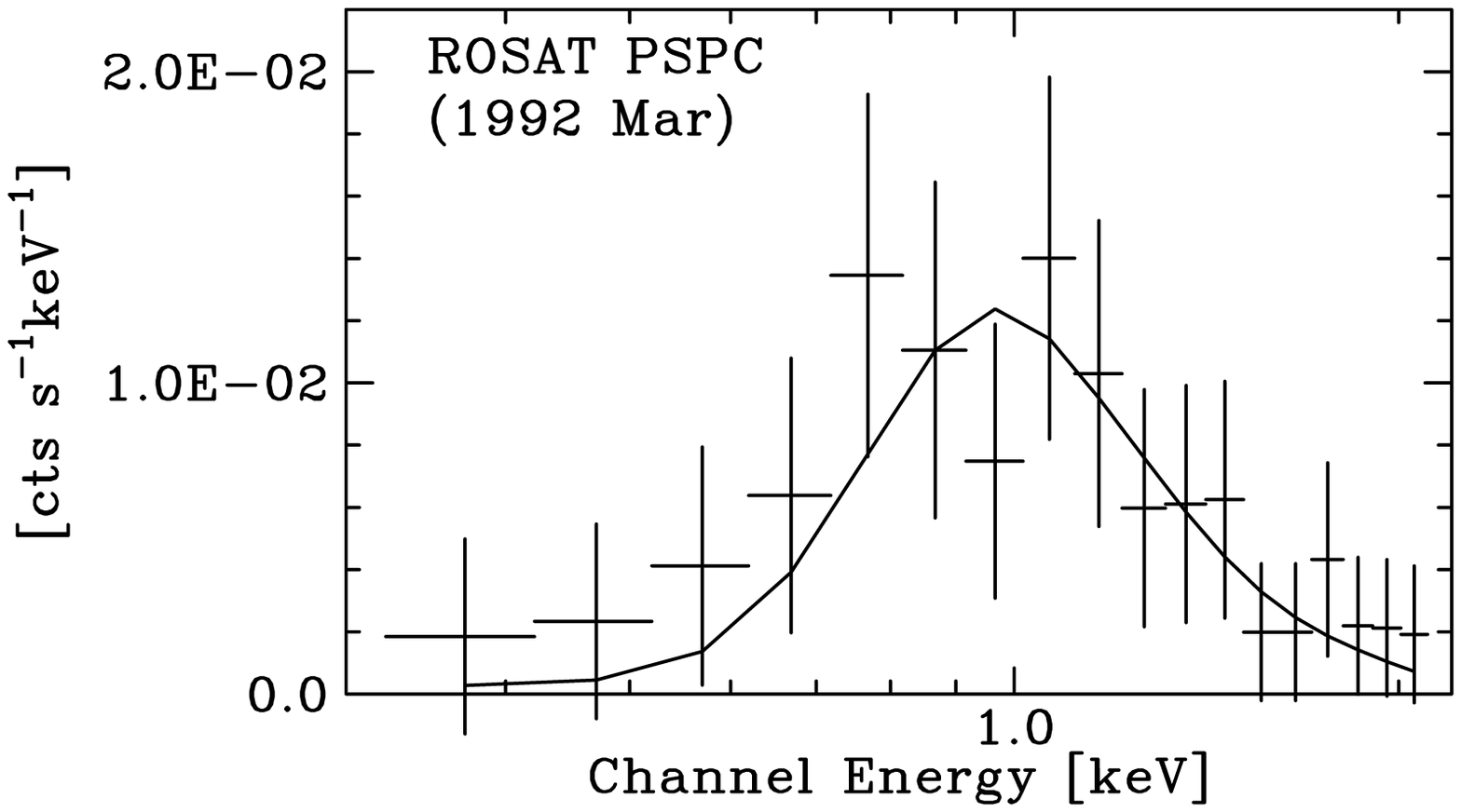}{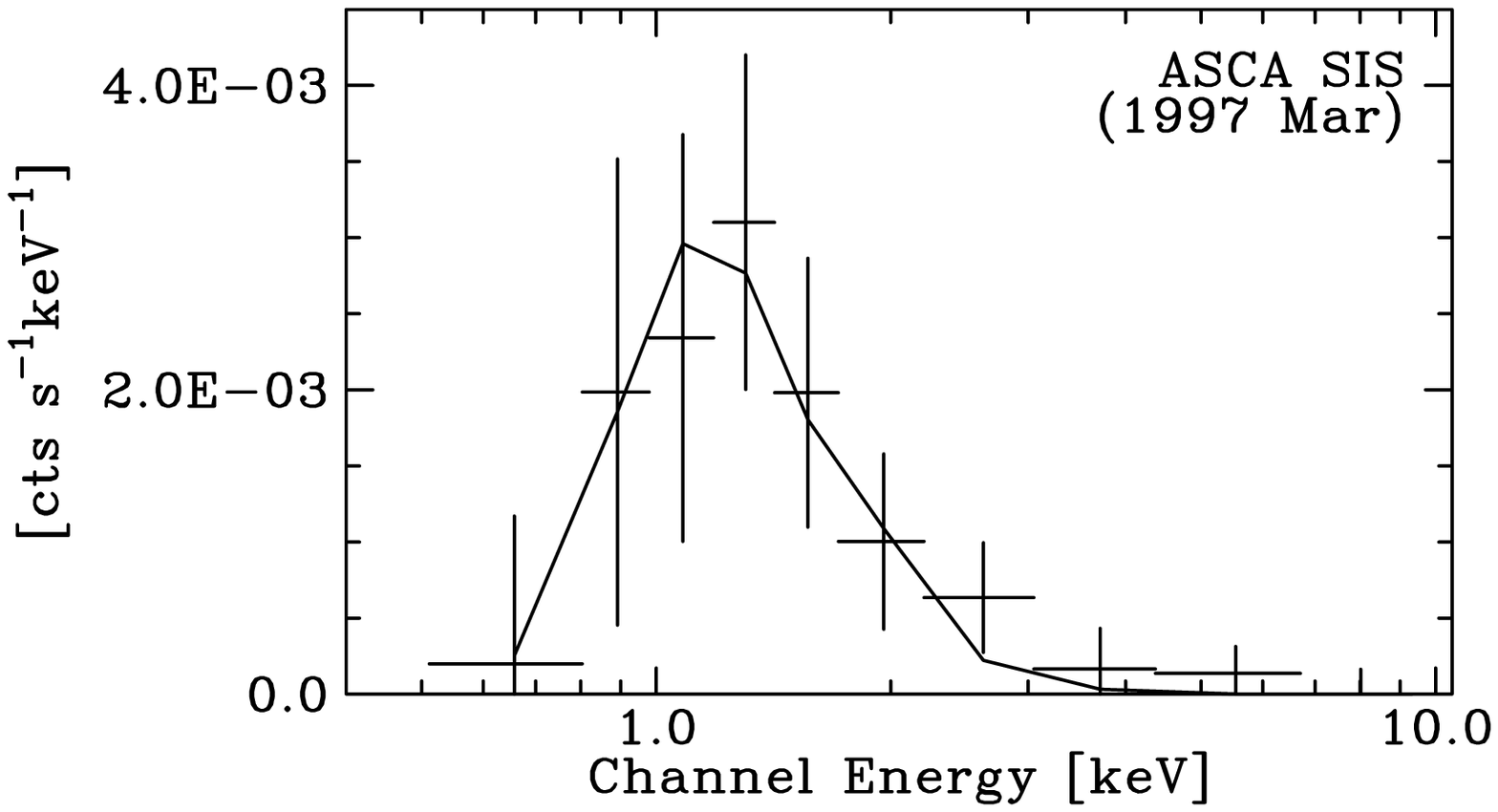}
\caption{(Left) The {\sl ROSAT\/} PSPC spectrum of RS~Oph from the
1992 Mar observation.  Data and the best-fit {\tt mekal} model are shown.
(Right) {\sl ASCA\/} SIS spectrum of RS~Oph, shown with the kT=0.6 keV
{\tt mekal} model.}
\end{figure}

Due to the marginal statistical quality of the data, spectral fitting
cannot provide tight constraints on the spectral shape.  However,
the same kT$\sim$0.6 keV plasma model that was used for the {\sl ROSAT\/}
spectrum provides a good description of the data below 2 keV, resulting
in a luminosity compatible with the first {\sl ROSAT\/} observation.
This model is shown with the SIS data in Figure~2 (right).  The possible
excess of data over this model above 2 keV may indicate that the emitting
plasma is hotter than 0.6 keV overall, or that there is an additional
contribution from a hotter plasma.

This is very different from the highly absorbed, bright X-ray component
seen in the quiescent recurrent nova, T~CrB \citep{Lea2007}, which is
otherwise very similar to RS~Oph.  T~CrB has a very hard, highly absorbed
X-ray emission with little flux below 2 keV, with an observed 2--10 keV
luminosity of 1--2 $\times 10^{33}$ ergs\,s$^{-1}$.  A 2--10 keV GIS
count rate of 1.2$\times 10^{-3}$ cts\,s$^{-1}$ corresponds to
an absorbed luminosity of 2--5$\times 10^{31}$ ergs\,s$^{-1}$ for
a kT=30 keV Bremsstrahlung spectrum absorbed with N$_{\rm H}$
of 10$^{22}$--10$^{24}$ cm$^{-2}$.  The unabsorbed luminosity may
be up to a factor of 10 higher for the extreme absorption case.

\section{Discussion}

Compared with the expected accretion luminosity from the boundary
layer, both the observed soft X-ray luminosity and the undetected
hard X-ray luminosity of RS~Oph are extremely low.  Thus, we have
a case of ``missing boundary layer'' problem in RS~Oph that is
perhaps more extreme than any seen in non-magnetic cataclysmic
variables (CVs; \citealt{vTea1996}).  In fact, we may not have detected
any boundary layer emission from RS~Oph at all: the properties
of the detected soft emission is quite similar to those seen
in several other symbiotic systems \citep{Mea1997}, for which
a colliding winds interpretation has been put forth.

It is unlikely that absorption alone can extinguish the hard X-ray
flux sufficiently.  The above analysis shows that even an extreme
N$_{\rm H}$ of 10$^{24}$ cm$^{-2}$ is insufficient in itself to reduce
a 2--10 keV X-ray component at the T~CrB level to below {\sl ASCA\/}
detectability level.  Moreover, such a high column density is
significantly above those seen in the hard X-ray component of T~CrB,
the X-rays from RS~Oph in outburst (up to 5$\times 10^{22}$ cm$^{-2}$
for the early shock emission; \citealt{Sea2006,Bea2006}), or in
optical data for RS~Oph is quiescence
($\sim$10$^{23}$ cm$^{-2}$; \citealt{AM1999}).

If the boundary layer is completely optically thick, hard X-rays
will not result.  Note that the resulting soft X-ray component is likely
to be hot in this case.  For example, a 1.4 M$_\odot$ white dwarf accreting at
$\sim$10$^{-8}$ M$_\odot$\,yr$^{-1}$ over 10\% of its surface area
should emit a kT$\sim$50 eV blackbody-like soft component,
which can only be hidden by a high, though realistic, intrinsic column.
In any case, the luck of a hard component is puzzling since, empirically
speaking, the boundary layer always appear to have a surface layer
optically thin enough to produce hard X-rays \citep{PR1985a}.  

If the white dwarf in RS~Oph is spinning rapidly, near the break-up
frequency, while that in T~CrB is not, this would explain the low
boundary layer luminosity in the former.  While UV spectroscopy of
non-magnetic CVs show that they harbor slowly rotating white dwarfs
\citep{S1999}, accretion of a few tenths of a solar mass should be
enough to spin up the CV primaries to near break-up.  CV primaries
are slow rotators probably because of the angular momentum removed
during nova outbursts, which
eject more mass than accreted \citep{Kea1991} and have an
extended common envelope phase \citep{LP1998}.  Neither is thought
to be the case with RS~Oph.  The high white dwarf mass inferred for RS~Oph
almost certainly means that it has accreted a significant amount
of mass, and therefore angular momentum.  If indeed the nova
mechanism is ineffective in removing angular momentum from the
white dwarf in RS~Oph, it is quite possibly a fast rotator.

However attractive this possibility may be, it remains a pure
conjecture.  Independent studies of this using, e.g., the morphology
of the ejecta \citep{OBea2006}, as well as future quiescent X-ray
observations, are strongly encouraged.

\acknowledgements 
This research has made use of data obtained from the High Energy Astrophysics
Science Archive Research Center (HEASARC), provided by NASA's Goddard Space
Flight Center.


\end{document}